\documentclass{moriond}

\def\be{\begin{equation}}
\def\ee{\end{equation}}
\def\bea{\begin{eqnarray}}
\def\eea{\end{eqnarray}}

\usepackage{xspace}
\usepackage{orcidlink}

\newcommand{\Photo}{}
\newcommand{\hu}{\xspace \ensuremath{{\rm km \, s^{-1} \, Mpc^{-1}}\xspace}}

\newcommand{\Msol}{\xspace\rm{M}_{\odot}\xspace}

\newcommand{\de}{{\rm d}}

\begin{document}
\vspace*{4cm}
\title{Gravitational Wave Cosmology: Be Careful of the Black Hole Mass Spectrum}

\author{ Grégoire Pierra \orcidlink{0000-0003-3970-7970} }

\address{Université Lyon, Université Claude Bernard Lyon 1, CNRS, IP2I Lyon/IN2P3,UMR 5822, F-69622 Villeurbanne, France}

\maketitle
\abstracts{Gravitational waves (GWs) from compact binary coalescences (CBCs) offer insights into the universe's expansion. The ``\textit{spectral siren}'' method, used without electromagnetic counterparts (EMC), infers cosmic expansion (Hubble constant $H_0$) by relating detector and source frame masses of black hole (BH) mergers. However, heuristic mass models (broken power law, power law plus peak, multipeak) may introduce biases in $H_0$ estimation, potentially up to 3$\sigma$ with 2000 detected GW mergers. These biases stem from models' inability to consider redshift evolution and unexpected mass features. Future GW cosmology studies should employ adaptable source mass models to address these issues.}

\section{Introduction\label{section: introduction}}
Since September 2015, the LIGO-Virgo-KAGRA collaboration has detected over 90 GW signals from CBCs \cite{KAGRA:2021vkt,LIGOScientific:2020ibl,LIGOScientific:2021usb}, offering direct measurements of luminosity distances ($d_L$) without relying on specific cosmological models. However, without an EMC, determining the redshift of the source to measure the universe's expansion rate becomes a complex task, a topic currently debated \cite{Freedman:2021ahq}. To address this, methods like the spectral siren technique have emerged, estimating redshift indirectly using the intrinsic relationship between redshift and masses in the detector and source frames \cite{Ezquiaga:2022zkx}. The accuracy of this approach relies on chosen phenomenological models for BBH source mass distribution. This study investigates how well common source mass models fit complex BBH distributions, aiming to uncover potential biases in $H_0$. Using synthetically generated BBH populations and hierarchical Bayesian inference \cite{Mastrogiovanni:2023zbw}, we simultaneously measure $H_0$ alongside phenomenological BBH population mass model.

\section{Method\label{section: method}}
\subsection{Simulation of GW observations\label{subsection: GW observations}}
The workflow for generating GW observations involves calculating an approximate optimal signal-to-noise ratio (SNR) from a BBH merger population, characterized by source masses and redshift. Instead of using the full waveform and matched filter, we use a proxy formula for the optimal SNR \cite{Fishbach:2018edt,Mastrogiovanni:2021wsd}:
\begin{equation}
       \rho = \delta \times \rho_{9} \left[\frac{\mathcal{M}_c}{\mathcal{M}_{c,{\rm 9}}} \right]^{\frac{5}{6}}\left[ \frac{d_{L,{\rm 9}}}{d_L}\right],
       \label{eq:SNR}
\end{equation}
where $\mathcal{M}_c$ is the chirp mass, $d_L$ is the luminosity distance, $\rho$ is the optimal SNR, and $\delta$ is a projection factor. We assume a flat $\Lambda CDM$ cosmology with $H_0 = 67.8$ $\hu$ and $\Omega_{m,0}=0.308$, based on Planck measurements \cite{Planck:2015fie}. The reference chirp mass, luminosity distance and optimal SNR are set to $\mathcal{M}_{c, {\rm 9}} = 25$ $\Msol$, $d_{L, {\rm 9}} = 1.5$ $\rm Gpc$ and $\rho_{9}=0$ \cite{KAGRA:2013rdx}. The projection factor $\delta$ accounts for the 3-detector network's sensitivity, including GW polarization and sky localization. To simulate detector noise, we calculate a "detected" SNR $\rho^{\rm det}$ from the optimal SNR using a $\chi^{2}$ distribution. A GW is considered detected if $\rho^{\rm det}$ exceeds 12 and the GW frequency from the innermost stable circular orbit (ISCO) is above 15 Hz. For this study, we assume perfect measurement of detector frame masses and luminosity distances to reduce computational load and focus on systematic effects from phenomenological mass models.

\subsection{Bayesian inference\label{subsection: bayesian inference}}
The detection of GW events is modeled as an inhomogeneous Poisson process with selection biases \cite{Vitale:2020aaz,Mastrogiovanni:2023zbw,Mastrogiovanni:2023emh,Mastrogiovanni:2021wsd}. For $N_{\rm GW}$ detected GW signals over an observation time $T_{\rm obs}$, the probability of obtaining a specific GW dataset \{x\} given population hyperparameters $\Lambda$ is:
\begin{eqnarray}
       \mathcal{L}(\{x\}|\Lambda) &\propto & e^{-N_{ \rm exp}(\Lambda)} \prod_i^{N_{\rm GW}} T_{\rm obs} \int \de \theta \de z  \mathcal{L}_{\rm GW}(x_i|\theta,z,\Lambda) \frac{1}{1+z} \frac{\de \rm N_{\rm CBC}}{\de \theta \de z \de t_s}(\Lambda).
       \label{eq}
\end{eqnarray}
The hyperparameters $\Lambda$ include BBH population parameters and cosmological expansion parameters (e.g., $H_0$ and $\Omega_{\rm m}$). The individual GW likelihood $\mathcal{L}_{\rm GW}(x_i|\theta,\Lambda)$ accounts for errors in intrinsic parameters $\theta$, specifically source frame masses. The CBC rate, parameterized for the spectral siren method, is:
\begin{equation}
       \frac{\de \rm N_{\rm CBC}}{\de \theta \de z \de t_s} = R_0 \psi(z;\Lambda) p_{\rm pop} (\vec{m}_{\rm s}|\Lambda) \times\frac{\de V_c}{\de z },
       \label{eq rate full}
\end{equation}
where $R_{0}$ is the local merger rate, $\vec{m}_{\rm s} = (\rm m^{\rm s}_{1},m^{\rm s}_{2})$ are the source frame masses, and $\psi(z;\Lambda)$ is the merger rate evolution with redshift. The term $\rm p_{\rm pop} (\vec{m}_{s}|\Lambda)$ is the probability density function for the source frame masses and $\frac{\de V_c}{\de z }$ is the differential of the comoving volume. We assume the CBC mass spectrum does not evolve with redshift, a common assumption in the literature \cite{LIGOScientific:2021aug,Gray:2023wgj,Mastrogiovanni:2023zbw,Mastrogiovanni:2023emh}. The implications of this assumption on $H_0$ bias are discussed subsequently.

\subsection{Population models of BBHs\label{subsection: population models}}
Phenomenological models for CBC merger rates, masses, and redshift play a pivotal role in this study. Current spectral siren analyses, exemplified by \cite{LIGOScientific:2021aug}, typically employ simple parametric models for the distribution of source frame masses and CBC merger rates. We utilize three mass models: Broken Power Law (BPL), Power Law plus Peak (PLP), and Multipeak (MLTP). The BPL \cite{LIGOScientific:2020kqk} is the most straightforward, featuring a power law with low mass smoothing and a break at $\rm m_{\rm break}$, accounting for effects of stellar progenitor metallicity and the pair-instability supernovae gap. The PLP \cite{Talbot:2018cva} expands on the BPL by adding a Gaussian peak to capture a potential BBH accumulation before the pair instability supernovae gap (PISN), with parameters like peak position and width allowing for flexibility. The MLTP model further extends the PLP by introducing a second Gaussian peak at higher masses, accommodating BBHs from second-generation black holes due to hierarchical mergers. 

\section{Results\label{section: results}}
\subsection{Sanity checks\label{subsection: sanity checks}}
\begin{figure*}[ht]
       \centering
       \includegraphics[width=\textwidth]{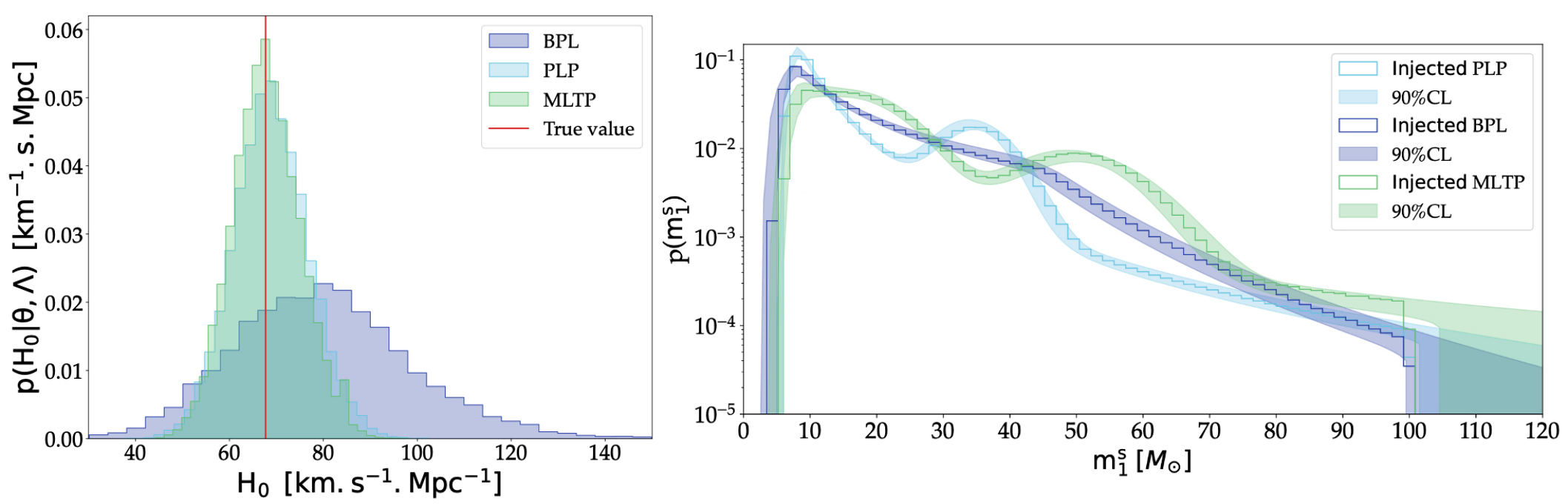}
       \caption{\textbf{Left:} Marginal posteriors of $H_0$ obtained via spectral sirens inference. The red line is the true value injected. \textbf{Right:}Reconstructed spectrum of the primary mass in source frame from the three inferences.}
       \label{fig: sanity check}
\end{figure*}

We simulate three test populations of BBH mergers, each with different source mass distributions and CBC merger rate. For each population, we generate 2000 detected GW events for use in spectral siren analysis. This number of detections typically ensures a data-informed posterior within the prior range $H_0 \in [20,140] \hu$. We employ hierarchical Bayesian inference to jointly estimate the cosmological parameter $H_0$ alongside all mass and CBC merger rate parameters. Left plot of Fig.~\ref{fig: sanity check} presents marginal posterior distributions for $H_0$ obtained for each source mass model. The true $H_0$ value is estimated within the 90\% C.I. for all simulations, with true values of other population parameters also contained in their respective 90\% C.I. The PLP and MLTP mass models yield more precise $H_0$ constraints than the BPL, owing to sharper features in the source mass spectrum.
Right plot of Fig.~\ref{fig: sanity check} displays posterior predictive checks (PPCs) for the BPL, PLP, and MLTP mass models. These PPCs demonstrate accurate reconstruction of mass spectrum structures by the inference, including all features present in the mass spectra. Thus, when using the correct mass model for inference, both population and cosmological expansion parameters are accurately recovered. Additionally, this test demonstrates the importance of features in the BBH mass spectrum for improving precision in cosmological expansion parameters.

\subsection{Linear redshift evolution\label{subsection: redshift evolution}}
\begin{figure*}[ht]
       \centering
       \includegraphics[width=\textwidth]{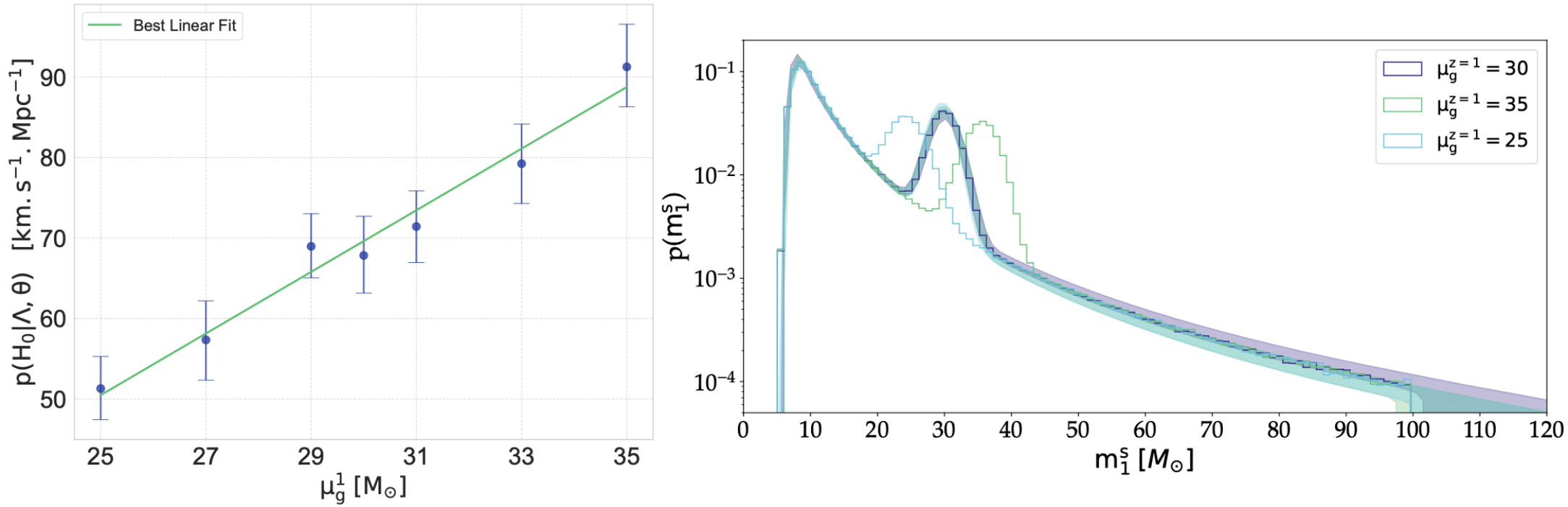}
       \caption{\textbf{Left:} Inferred value of $H_0$ as a function of the position of the Gaussian peak at $z=1$. \textbf{Right:}Reconstructed spectrum of the primary mass in source frame from the three inferences.}
       \label{fig: linear_z_evol}
\end{figure*}

We conduct a test to assess the robustness of redshift-independent mass models to a potential evolution of the mass spectrum with redshift. We assume a linear dependence between mass spectrum features and redshift. Here, we modify the PLP mass model to include a linearly evolving peak position with redshift, while keeping the edges fixed. This evolution could arise from interplay between PISN mass scale and binary formation-to-merger time-delays \cite{Mukherjee:2021rtw,2023A&A...677A.124K}. The modified PLP model incorporates redshift evolution as follows:
\begin{equation}
       \mu_{g}(z) = \mu_{g}^{0} + z \left(\mu_{g}^{1} -\mu_{g}^{0} \right),
       \label{eq:peak evolution in z}
\end{equation}
where $\mu_{g}^{0}$ and $\mu_{g}^{1}$ denote peak positions at $z=0$ and $z=1$ respectively, with $\mu_{g}^{0} = 30\Msol$ and varying $\mu_{g}^{1}$ from $25 \Msol$ to $35 \Msol$. Each simulation involves hierarchical Bayesian inference using the redshift-independent PLP to reconstruct $H_0$ and other population parameters from 2000 GW detections. Left of Fig.~\ref{fig: linear_z_evol} shows $H_0$ estimations with 68.3\% C.I. for these analyses, indicating $H_0$ variations as $\mu_{g}^{1}$ changes. Even a mild mass spectrum evolution biases the inferred $H_0$ significantly, linearly proportional to the redshift evolution magnitude. This bias arises from inaccurate reconstruction of the source-frame mass spectrum, consistent across independent simulation realizations. Right of Fig.~\ref{fig: linear_z_evol} presents PPC for this test. The PLP model fails to capture this evolution, consistently reconstructing a peak at 30 $\Msol$. Misplacement of the Gaussian component at higher redshifts induces $H_0$ bias: for $\mu_{\rm g}^{1}>\mu_{\rm g}^{0}$, $H_0$ is biased higher, and vice versa. This suggests potential bias in GW-based $H_0$ estimation from redshift-evolving mass spectrum features.

\subsection{Complex BBHs population\label{subsection: complex population}}
\begin{figure*}[ht]
       \centering
       \includegraphics[width=\textwidth]{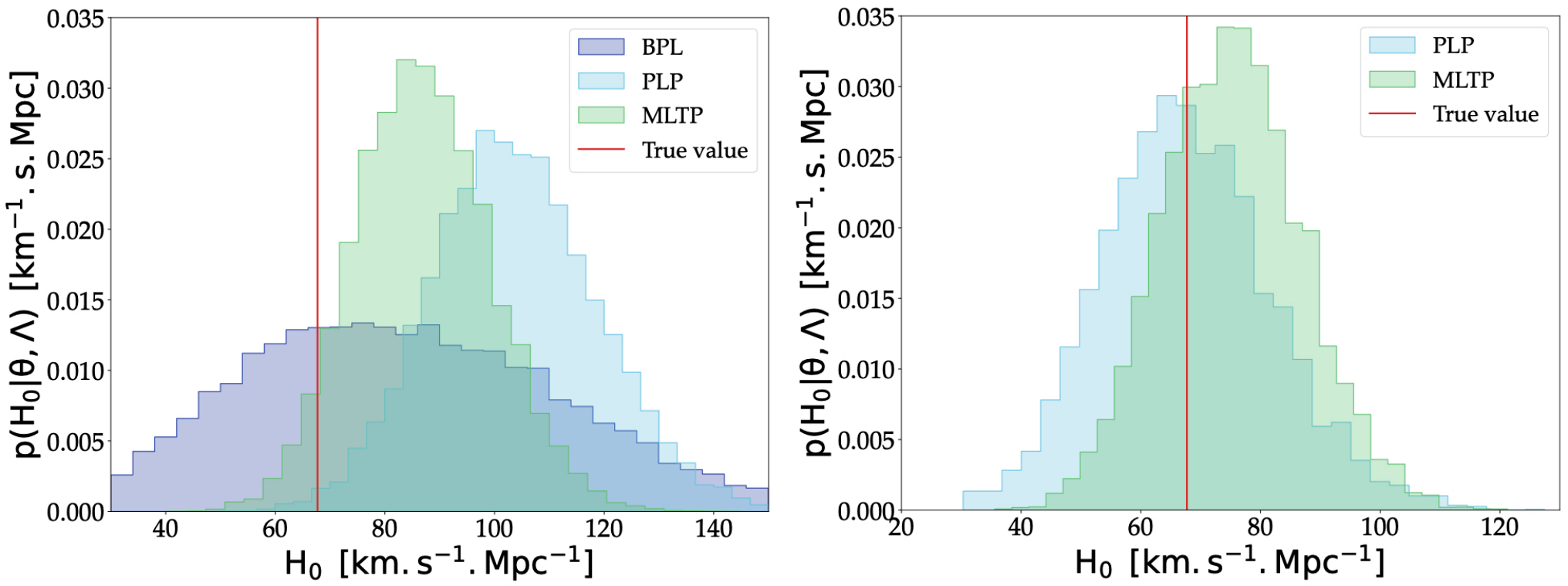}
       \caption{\textbf{Left:} Marginal posterior of $H_0$ with the A03 BBH population. \textbf{Right:}Marginal posterior of $H_0$ with the A03 BBH population, where the redshift evolution has been blinded to the masses.}
       \label{fig: complex_model}
\end{figure*}

Utilizing an astrophysically motivated synthetic BBH catalog (A03) \cite{Mapelli:2021gyv}, we generate GW catalogs containing $2000$ detected events. Each catalog undergoes full hierarchical Bayesian inference to estimate posteriors on $H_0$ and other population parameters, using redshift rate and three redshift-independent BBH mass models. Left of Fig.~\ref{fig: complex_model} depicts the marginal posterior distributions of estimated Hubble constants. Both PLP and MLTP mass models exhibit an $H_0$ bias towards higher values. Specifically, PLP yields $H_0 = 103^{+11}_{-10} \hu$ and $H_0 = 87^{+9}_{-8} \hu$ at 68.3\% C.I., excluding the true value of $H_0$ from the 99.7\% and 95\% C.I. respectively. In contrast, the BPL mass model yields an uninformative $H_0$ posterior, encompassing the injected $H_0$ within the 68\% C.I. The lack of strong mass features in BPL contributes to this less informative posterior. PLP and MLTP underestimate the presence of BBHs between $40\Msol$ and $80 \Msol$, leading to an overestimation of $H_0$ by pushing events to higher redshifts with lower source masses. Although PLP and MLTP possess enough flexibility to approximate smooth mass distributions akin to the A03 catalog, they fail to accurately reconstruct the mass spectrum and consequently introduce $H_0$ bias.
To further explore the potential impact of redshift evolution on the mass spectrum and its influence on $H_0$ bias, we conducted a simulation by blinding the A03 mass spectra to redshift evolution. This involved randomly shuffling pairs of BBH merger redshifts and masses, removing the redshift dependency while preserving the mass spectrum's shape. We focused on the PLP and MLTP mass models due to their notable $H_0$ bias. Right of Fig.~\ref{fig: complex_model} presents the joint inference results with the PLP mass model, yielding $H_0 = 67.0^{+9.7}_{-8.9} \hu$. Similar results were obtained with MLTP ($H_0 = 75.13^{+7.8}_{-7.9} \hu$), with the true $H_0$ value falling within the 68.3\% C.I. of the posterior. Our test confirms that redshift evolution of the mass spectrum is the significant contributor to $H_0$ bias.

\section{Conclusion\label{section: conclusion}}
This study delves into systematic biases affecting $H_0$ estimation and BBH mass spectrum reconstruction in spectral siren Cosmology. Through simulations with various mass models and synthetic BBH catalogs, we uncovered critical insights. We revealed that simplistic models lacking sharp features may bias $H_0$ if true BBH populations exhibit local over-densities. Additionally, neglecting redshift evolution in source frame mass features can skew $H_0$ estimates, as underestimated mass spectra shift $H_0$ higher, and vice versa. Using the A03 BBH catalog, we found $H_0$ biases of up to $3\sigma$ for PLP and MLTP inferences. Removing redshift evolution eliminated biases. Future GW cosmology studies should employ flexible mass models capable of accommodating such complexities to avoid biased $H_0$ estimations, especially with numerous detected GW events.

\section*{Acknowledgments}
The authors are grateful for computational resources provided by the LIGO Laboratory and supported by the National Science Foundation Grants PHY-0757058 and PHY-0823459. We thank Viola Sordini for the helpful comments and discussions on the manuscript.  MM acknowledges financial support from the European Research Council for the ERC Consolidator grant DEMOBLACK, under contract no. 770017, and from the German Excellence Strategy via the Heidelberg Cluster of Excellence (EXC 2181 - 390900948) STRUCTURES.

\bibliographystyle{unsrt}    
\bibliography{bibliography}

\end{document}